# Analytical and semi-analytical solutions to the kinetic equation with Coulomb collision term and a monoenergetic source function

**P R Goncharov**

Saint-Petersburg Polytechnic University, 195251, Russia
and RRC "Kurchatov Institute", Moscow, 123182, Russia

E-mail: goncharov@phtf.stu.neva.ru

**Abstract.** Complete and physically adequate analytical and semi-analytical solutions have been obtained using a practical dimensionless form of kinetic equation assuming azimuthal symmetry and Maxwellian distributions of target plasma species. Formerly considered simplified equations with truncated Coulomb collision term do not conserve the number of particles, are inapplicable to describe high energy distribution tails, and are also essentially unable to demonstrate the Maxwellization process naturally observed in the low energy region of correct distributions. The results may be useful in numerical modeling and in experimental data analysis, especially concerning nuclear processes and advanced localized, angle-resolved suprathermal particle diagnostics.

## 1. Introduction

Let $f_\alpha(\mathbf{v})$ be the sought velocity distribution function of test particles of type $\alpha$, and $f_\beta(\mathbf{v}')$ be the known velocity distribution functions of target plasma species counted by index $\beta$. Functions $f_\alpha(\mathbf{v})$ and $f_\beta(\mathbf{v}')$ are normalized to unity. To obtain a practical form of the equation to be solved, we use Coulomb collision term [1] expressed via partial potential functions [2,3] corresponding to the interaction of test particles $\alpha$ with particular species $\beta$ of target plasma

$$\Phi_\beta = -\frac{1}{4\pi}\int \frac{n_\beta f_\beta(\mathbf{v}')}{u} d^3\mathbf{v}', \tag{1}$$

$$\Psi_\beta = -\frac{1}{8\pi}\int u n_\beta f_\beta(\mathbf{v}') d^3\mathbf{v}', \tag{2}$$

where $n_\beta$ denotes the density of particles $\beta$, and $u = |\mathbf{v}-\mathbf{v}'|$ denotes the magnitude of the relative velocity of particles $\alpha$ and $\beta$. As shown in [4], assumptions of azimuthal symmetry (i.e. $\frac{\partial}{\partial\varphi}=0$) and angle isotropy of distribution functions $f_\beta(\mathbf{v}')$ of target plasma species lead to the following expression for the partial collision term in spherical polar coordinates in velocity space:

$$C_{\alpha\beta} = \frac{L_{\alpha\beta}}{v^2}\frac{\partial}{\partial v}\left(v^2\left(\frac{m_\alpha}{m_\beta}(n_\alpha f_\alpha)\frac{\partial \Phi_\beta}{\partial v} - \frac{\partial^2 \Psi_\beta}{\partial v^2}\frac{\partial(n_\alpha f_\alpha)}{\partial v}\right)\right) - \frac{L_{\alpha\beta}}{v\sin\vartheta}\frac{\partial}{\partial\vartheta}\left(\frac{\sin\vartheta}{v^2}\frac{\partial \Psi_\beta}{\partial v}\frac{\partial(n_\alpha f_\alpha)}{\partial\vartheta}\right), \quad (3)$$

where $m_\alpha$ and $m_\beta$ are the masses of particles of species $\alpha$ and $\beta$, respectively, $n_\alpha$ denotes the density of particles $\alpha$,

$$L_{\alpha\beta} = \frac{\left(4\pi Z_\alpha Z_\beta e^2\right)^2 \Lambda}{m_\alpha^2}, \quad (4)$$

$Z_\alpha$ and $Z_\beta$ are the electric charge numbers of particles of species $\alpha$ and $\beta$, respectively, $e$ is the elementary charge, $\Lambda$ is Coulomb logarithm. The full collision term is

$$C_\alpha = \sum_\beta C_{\alpha\beta}. \quad (5)$$

## 2. Dimensionless notation

We follow the dimensionless approach of monograph [4], using a slightly different notation. Namely, we do not introduce the injection velocity into the expression for the collision term, since this is an external parameter, and it is more natural to retain it in the test particle source function only. Unlike [4], in our notation a small factor $(m_e/m_\alpha)^{1/3} < 0.1$, when species $\alpha$ are ions, appears naturally in the velocity diffusion term without introducing the ratio of the electron temperature to the test particle injection energy.

Defining generalized temperatures for all target plasma species

$$T_\beta = \frac{2}{3}\left\langle \frac{m_\beta v_\beta^2}{2} \right\rangle = \frac{2}{3}\int_0^{+\infty} \frac{m_\beta v_\beta^2}{2} f_\beta(v_\beta) 4\pi v_\beta^2 dv_\beta, \quad (6)$$

three partial (i.e. corresponding to the particular species $\beta$) dimensionless functions

$$a_\beta(v) = -\frac{4\pi m_\beta}{n_\beta T_\beta} v^3 \frac{\partial^2 \Psi_\beta}{\partial v^2}, \quad (7)$$

$$b_\beta(v) = \frac{4\pi}{n_\beta} v^2 \frac{\partial \Phi_\beta}{\partial v}, \quad (8)$$

$$c_\beta(v) = -\frac{4\pi}{n_\beta}\sqrt{\frac{2T_\beta}{m_\beta}}\frac{1}{v}\frac{\partial \Psi_\beta}{\partial v}, \quad (9)$$

and the dimensional constants $v_c$ [cm/s] and $\tau_s$ [s]

$$v_c^3 = \frac{m_e}{m_\alpha}\left(\frac{2T_e}{m_e}\right)^{3/2}, \tag{10}$$

$$\tau_s = \left(\frac{m_\alpha}{Z_\alpha e \omega_{pe}}\right)^2 \frac{v_c^3}{\Lambda m_e}, \tag{11}$$

where

$$\omega_{pe} = \sqrt{\frac{4\pi n_e e^2}{m_e}} \tag{12}$$

is the electron plasma frequency, and a dimensionless parameter

$$\varepsilon = \left(\frac{m_e}{m_\alpha}\right)^{1/3}, \tag{13}$$

we then write three dimensionless functions summed over all species $\beta$

$$a(v) = \varepsilon \frac{m_\alpha}{n_e} \sum_\beta \frac{n_\beta Z_\beta^2}{m_\beta} \frac{T_\beta}{T_e} a_\beta(v), \tag{14}$$

$$b(v) = \frac{m_\alpha}{n_e} \sum_\beta \frac{n_\beta Z_\beta^2}{m_\beta} b_\beta(v), \tag{15}$$

$$c(v) = \frac{v_c}{n_e} \sum_\beta \frac{n_\beta Z_\beta^2}{\sqrt{2T_\beta / m_\beta}} c_\beta(v), \tag{16}$$

and, finally, the collision term equivalent to (5) in the form

$$C_\alpha = \frac{v_c^3}{\tau_s} \frac{1}{v^2} \left( \frac{\partial}{\partial v} \left( v_c^2 \frac{a(v)}{2v} \frac{\partial (n_\alpha f_\alpha)}{\partial v} + b(v)(n_\alpha f_\alpha) \right) + \frac{c(v)}{v_c} \frac{1}{\sin\vartheta} \frac{\partial}{\partial \vartheta} \left( \sin\vartheta \frac{\partial (n_\alpha f_\alpha)}{\partial \vartheta} \right) \right). \tag{17}$$

For the particular case when all target plasma species are Maxwellian

$$f_\beta(v_\beta) = \left(\frac{m_\beta}{2\pi T_\beta}\right)^{3/2} e^{-\frac{m_\beta v_\beta^2}{2T_\beta}} \tag{18}$$

the derivatives $\frac{\partial \Phi_\beta}{\partial v}$, $\frac{\partial \Psi_\beta}{\partial v}$, and $\frac{\partial^2 \Psi_\beta}{\partial v^2}$ were calculated in [4], and functions (7)-(9) were expressed via Chandrasekhar function

$$G(z) = \frac{2}{\sqrt{\pi} z^2} \int_0^z x^2 e^{-x^2} dx = \frac{1}{2z^2} erf(z) - \frac{1}{\sqrt{\pi} z} e^{-z^2} \tag{19}$$

as follows:

$$a_\beta(v) = b_\beta(v) = 2\upsilon_\beta^2 G(\upsilon_\beta), \tag{20}$$

$$c_\beta(v) = \frac{1}{\sqrt{\pi}} e^{-\upsilon_\beta^2} + \left(\upsilon_\beta - \frac{1}{2\upsilon_\beta}\right) G(\upsilon_\beta), \tag{21}$$

where

$$\upsilon_\beta = v/v_{T_\beta}\ ,\quad v_{T_\beta} = \sqrt{2T_\beta/m_\beta}\ . \tag{22}$$

Note, that

$$a_\beta(v)\xrightarrow[v\to 0]{}0\ ,\quad a_\beta(v)\xrightarrow[v\to\infty]{}1\ ; \tag{23}$$

$$c_\beta(v)\xrightarrow[v\to 0]{}\frac{2}{3\sqrt{\pi}}\ ,\quad c_\beta(v)\xrightarrow[v\to\infty]{}0\ . \tag{24}$$

To compute (20), (21) in the vicinity of $v = 0$ it is useful to apply the decomposition given in [5]

$$erf(z) = \frac{2}{\sqrt{\pi}}e^{-z^2}\sum_{k=0}^{\infty}\frac{2^k z^{2k+1}}{(2k+1)!!} = \frac{2}{\sqrt{\pi}}e^{-z^2}\left(z + \frac{2}{3}z^3 + \frac{4}{15}z^5 + ...\right). \tag{25}$$

Function $b(v)$ is related to the dynamic friction force and is responsible for the slowing-down process. Functions $a(v)$ and $c(v)$ are both related to the diffusion tensor in velocity space. The term with $c(v)$ in (17) contains only the angle derivatives and is responsible for the pitch angle scattering. The term with $a(v)$ describes the velocity diffusion process. For an isothermal Maxwellian plasma $a(v) = \varepsilon b(v)$ , and $\varepsilon$ is a small parameter when the test particles $\alpha$ are significantly heavier than electrons, while for electrons $a(v) = b(v)$ .

To calculate the collision term (17) for velocities much greater than thermal velocities of target plasma ions $v_{T_i} = \sqrt{2T_i/m_i}$ and much smaller than the thermal velocity of target plasma electrons $v_{T_e} = \sqrt{2T_e/m_e}$ , i.e. for $v_{T_i} \ll v \ll v_{T_e}$ , the following simplified formulas may be used instead of applying (14)-(16) and (20), (21):

$$a(v) = \varepsilon\left(Z^{(a)} + \frac{m_\alpha}{m_e}\frac{4}{3\sqrt{\pi}}\left(v/v_{T_e}\right)^3\right), \tag{26}$$

$$b(v) = Z^{(b)} + \frac{m_\alpha}{m_e}\frac{4}{3\sqrt{\pi}}\left(v/v_{T_e}\right)^3 , \tag{27}$$

$$c(v) = Z^{eff}\frac{v_c}{2v} + \frac{2}{3\sqrt{\pi}}\frac{v_c}{v_{T_e}} , \tag{28}$$

where

$$Z^{(a)} = \frac{m_\alpha}{n_e T_e}\sum_i \frac{Z_i^2 n_i T_i}{m_i} , \tag{29}$$

$$Z^{(b)} = \frac{m_\alpha}{n_e}\sum_i \frac{Z_i^2 n_i}{m_i} , \tag{30}$$

$$Z^{eff} = \frac{1}{n_e}\sum_i Z_i^2 n_i , \tag{31}$$

and the summation in (29)-(31) is over all ion species of the target plasma. The first terms in right-hand sides of expressions (26)-(28) represent the contributions of target plasma ions, and the second terms represent the contributions of target plasma electrons. Ion and electron contributions to the simplified slowing-down term governed by (27) are equal when

$$v = \left(3\sqrt{\pi} Z^{(b)}/4\right)^{1/3} v_c , \tag{32}$$

therefore, (32) is often called a 'critical velocity'.

## 3. Working form of the equation

Consider Boltzmann kinetic equation for the sought distribution function $n_\alpha f_\alpha$, neglecting the spatial inhomogeneity and the electric field

$$\frac{\partial \left(n_\alpha f_\alpha\right)}{\partial t} = C_\alpha + S_\alpha , \tag{33}$$

where $C_\alpha$ is the collision term corresponding to collisions of particles $\alpha$, originating from a monoenergetic beam in a magnetically confined plasma, with particles of all species of the target plasma, and $S_\alpha$ is the source function of particles $\alpha$. The collision term $C_\alpha$ calculated by (14)-(17) and (20), (21) is as exact as [1] with only two assumptions, viz., that the azimuthal symmetry takes place and that the target plasma is Maxwellian. The azimuthal symmetry is a reasonable assumption since Larmor gyration tends to average-out the angle $\varphi$ dependence, and the distribution function $f_\alpha(\mathbf{v})$ in a strong magnetic field is axially symmetric, i.e. it is a function of the velocity magnitude $v$ and the pitch angle $\vartheta$.

Simplified equations solved in [6,7] correspond to $b(v)$ given by (27), and $c(v) = Z^{eff} v_c / 2v$, i.e. the first term in (28), while velocity diffusion term with $a(v)$ is incorrect in both [6] and [7]. In case of isothermal Maxwellian target plasma, i.e. $T_\beta = T \ \ \forall \beta$, the correct Coulomb collision operator applied to the Maxwellian distribution function with the equilibrium temperature $T_\alpha = T$ results in nullification of the collision term. As opposed to [6,7], this fundamental physical property preserves if we use the correct expressions for dimensionless functions $a(v)$, $b(v)$ and $c(v)$ given in section 2. The purpose of the subsequent sections is to obtain the exact and physically adequate solutions of (33) without simplifications.

Thus, the working form of equation (33) is

$$\frac{\partial\phi}{\partial\tau}=\frac{a(u)}{2u^3}\frac{\partial^2\phi}{\partial u^2}+\left(\frac{b(u)}{u^2}-\frac{a(u)}{2u^4}+\frac{1}{2u^3}\frac{\partial a}{\partial u}\right)\frac{\partial\phi}{\partial u}+\frac{1}{u^2}\frac{\partial b}{\partial u}\phi+\frac{c(u)}{u^2}\frac{\partial}{\partial\zeta}\left(1-\zeta^2\right)\frac{\partial\phi}{\partial\zeta}+\tau_s S_\alpha(u,\zeta,\tau)\,, \tag{34}$$

where $\phi(u,\zeta,\tau)\equiv n_\alpha f_\alpha$ is the sought function, $u=v/v_c$ is the dimensionless velocity, $\tau=t/\tau_s$ is the dimensionless time, and $\zeta=\cos\vartheta$ is the pitch angle cosine. The stationary monoenergetic isotropic source function is

$$S_\alpha(u)=\frac{S_0}{4\pi v_c^3}\frac{1}{u^2}\delta(u-u_0)\,, \tag{35}$$

and the stationary monoenergetic anisotropic source function is

$$S_\alpha(u,\zeta)=\frac{S_0}{2\pi v_c^3}\frac{1}{u^2}\delta(u-u_0)\mathcal{Z}(\zeta)\,, \tag{36}$$

where $\delta(u-u_0)$ is delta-function, $u_0=v_0/v_c$ is the dimensionless injection velocity, and $\mathcal{Z}(\zeta)$ is the unity-normalized angle distribution of the source. The source function given by either (35) or (36) is normalized to the source rate $S_0$ [cm$^{-3}$s$^{-1}$], i.e. the number of particles of type $\alpha$ injected in unit volume in unit time,

$$\int S_\alpha d^3\mathbf{v}=2\pi v_c^3\int_0^\infty u^2du\int_{-1}^1 d\zeta S_\alpha(u,\zeta)=S_0\,. \tag{37}$$

## 4. Isotropic problem

### 4.1. Slowing-down

Analytical steady state solution of equation (34) taking into account only the dynamic friction force, but not the diffusion in velocity space, i.e. with $a(u)=0$, $c(u)=0$, and isotropic $S_\alpha(u)$ given by (35), can be obtained by variable separation method for the corresponding stationary homogeneous first order ordinary differential equation and then variation of constant. The resulting isotropic distribution

$$\phi(u)=\frac{S_0\tau_s}{4\pi v_c^3}\frac{1}{b(u)}H(u_0-u)\,, \tag{38}$$

where $H(u_0-u)$ is Heaviside step function, is typically used plugging the simplified formula (27) for $b(u)$ instead of (15) and (20). We reproduce this simple stationary slowing-down distribution similar to [8-10] here as a reference for comparison with our solutions below.

### 4.2. Slowing-down and velocity diffusion in isothermal plasma

Simplified solution (38), neglecting the diffusion in velocity space, is inherently unable to describe the Maxwellization process. It is also cutting off the high energy distribution tail and therefore is inapplicable at $u > u_0$. To obtain a physically adequate solution, let us first consider an isotropic problem assuming that $c(u) = 0$, $S_\alpha(u)$ is given by (35), and all target plasma species are in thermal equilibrium i.e. $T_\beta = T \quad \forall \beta$. Equation (34) with $c(u) = 0$ reduces to

$$p(u)\frac{\partial^2 \phi}{\partial u^2} + q(u)\frac{\partial \phi}{\partial u} + r(u)\phi(u) = f(u), \tag{39}$$

where

$$p(u) = \frac{a(u)}{2u^3}, \tag{40}$$

$$q(u) = \frac{b(u)}{u^2} - \frac{a(u)}{2u^4} + \frac{1}{2u^3}\frac{\partial a}{\partial u}, \tag{41}$$

$$r(u) = \frac{1}{u^2}\frac{\partial b}{\partial u}, \tag{42}$$

$$f(u) = -\frac{S_0 \tau_s}{4\pi v_c^3}\frac{1}{u^2}\delta\left(u - u_0\right). \tag{43}$$

As mentioned above, in this isothermal case $a(u) = \varepsilon b(u)$, and it can be easily checked by substitution that Maxwellian function

$$\phi_1(u) = e^{-u^2/\varepsilon} \tag{44}$$

is a partial solution of the homogeneous equation corresponding to (39). It can also be easily verified by substitution, that the second independent solution of the homogeneous equation is

$$\phi_2(u) = e^{-u^2/\varepsilon}\int_{u_l}^{u}\frac{\bar{u}e^{\bar{u}^2/\varepsilon}}{a(\bar{u})}d\bar{u}. \tag{45}$$

Now that $\phi_1(u)$ and $\phi_2(u)$ are determined, we can find the solution of the inhomogeneous equation (39) in the form

$$\phi(u) = C_1(u)\phi_1(u) + C_2(u)\phi_2(u), \tag{46}$$

using Lagrange method of variation of constants. To satisfy (39) we require that

$$\left.\begin{aligned} C_1'(u)\phi_1(u) + C_2'(u)\phi_2(u) &= 0 \\ C_1'(u)\phi_1'(u) + C_2'(u)\phi_2'(u) &= f(u)/p(u) \end{aligned}\right\}. \tag{47}$$

This is a system of linear algebraic equations with respect to $C_1'(u)$ and $C_2'(u)$. Its solution is

$$C_1'(u) = \frac{S_0\tau_s}{2\pi v_c^3}\delta(u-u_0)\int_{u_l}^{u}\frac{\bar{u}e^{\bar{u}^2/\varepsilon}}{a(\bar{u})}d\bar{u}\,, \tag{48}$$

$$C_2'(u) = -\frac{S_0\tau_s}{2\pi v_c^3}\delta(u-u_0)\,. \tag{49}$$

Integrating (48) and (49), we obtain

$$C_1(u) = \frac{S_0\tau_s}{2\pi v_c^3}H(u-u_0)\int_{u_l}^{u_0}\frac{\bar{u}e^{\bar{u}^2/\varepsilon}}{a(\bar{u})}d\bar{u} + K_1\,, \tag{50}$$

$$C_2(u) = \frac{S_0\tau_s}{2\pi v_c^3}H(u_0-u) + K_2\,, \tag{51}$$

where $K_1$ and $K_2$ are arbitrary constants.

Finally, the partial solution of the inhomogeneous equation (39) is

$$\phi_p(u) = \frac{S_0\tau_s}{2\pi v_c^3}H(u-u_0)e^{-u^2/\varepsilon}\int_{u_l}^{u_0}\frac{\bar{u}e^{\bar{u}^2/\varepsilon}}{a(\bar{u})}d\bar{u} + \frac{S_0\tau_s}{2\pi v_c^3}H(u_0-u)e^{-u^2/\varepsilon}\int_{u_l}^{u}\frac{\bar{u}e^{\bar{u}^2/\varepsilon}}{a(\bar{u})}d\bar{u}\,, \tag{52}$$

and the general solution of the homogeneous equation corresponding to (39) is

$$\phi_h(u) = K_1 e^{-u^2/\varepsilon} + K_2 e^{-u^2/\varepsilon}\int_{u_l}^{u}\frac{\bar{u}e^{\bar{u}^2/\varepsilon}}{a(\bar{u})}d\bar{u}\,. \tag{53}$$

The general solution of (39) is

$$\phi(u) = \phi_h(u) + \phi_p(u)\,. \tag{54}$$

Two other independent equations are required to find the constants $K_1$ and $K_2$. A reasonable condition to determine $K_2$ is that $\phi(u)\xrightarrow[u\to\infty]{}0$, therefore, $K_2 = 0$. There is no particular boundary condition at $u = 0$. The meaning of the multiplier in the Maxwellian term $K_1 e^{-u^2/\varepsilon}$ can be explained using the normalization condition. Since our distribution function is normalized to the number of particles, the integral over the entire velocity space should be equal to the density of particles of type $\alpha$, which, in turn, equals the source rate $S_0$ times the duration $\kappa\tau_s$ of source action, i.e.

$$4\pi v_c^3\int_0^\infty u^2\phi(u)du = \kappa\tau_s S_0\,, \tag{55}$$

where $\kappa$ is a dimensionless constant, and $\kappa\tau_s$ is the time required to attain the steady state. Using the fact that

$$\int_0^\infty u^2 e^{-u^2/\varepsilon}du = \frac{\sqrt{\pi}}{4}\varepsilon^{3/2}\,, \tag{56}$$

we obtain the relationship between $K_1$ and $\kappa$

$$K_1 = \frac{4}{\sqrt{\pi}\varepsilon^{3/2}}\left(\frac{\kappa S_0 \tau_s}{4\pi v_c^3} - \int_0^\infty u^2 \phi_p(u)du\right). \tag{57}$$

### 4.3. Slowing-down and velocity diffusion in nonisothermal plasma

It is more difficult to obtain the exact analytical solution, when target plasma species have different temperatures. In this subsection we describe a numerical solution of the isotropic problem (39)-(42). For the numerical treatment instead of $\delta(u-u_0)$ we use a delta-like function

$$\mathcal{D}(u-u_0) = \frac{1}{\Delta\sqrt{\pi}} e^{-(u-u_0)^2/\Delta^2}, \tag{58}$$

where $\Delta$ is a small dimensionless parameter corresponding the peak width, and the source function $S_\alpha(u)$ given by

$$S_\alpha(u) = \frac{S_0}{4\pi v_c^3}\frac{1}{u^2}\mathcal{D}(u-u_0). \tag{59}$$

The right-hand side of (39) is then

$$f(u) = -\frac{S_0 \tau_s}{4\pi v_c^3}\frac{1}{u^2}\frac{1}{\Delta\sqrt{\pi}} e^{-(u-u_0)^2/\Delta^2}. \tag{60}$$

Note that $a(u)$ and $b(u)$ are different functions given by (14), (15), and (20), and there is no simple proportionality between them in contrast to the isothermal case considered in subsection 4.2.

To solve the problem formulated by (39)-(42), and (60) over the interval $[u_L, u_R]$ we introduce a uniform grid

$$u_k = u_L + (k-1)h, \tag{61}$$

where $k \in \overline{1,N}$,

$$h = \frac{u_R - u_L}{N-1}, \tag{62}$$

and $N$ is the grid dimension. Using forward difference derivatives at $u = u_1 = u_L$, i.e. for $k=1$, central difference derivatives at the inner grid points, i.e. for $k \in \overline{2,(N-1)}$, and backward difference derivatives at $u = u_N = u_R$, i.e. $k = N$, we approximate equation (39) by a system of linear algebraic equations

$$\mathbf{A}\boldsymbol{\phi} = \mathbf{f}, \tag{63}$$

where $\boldsymbol{\phi} = (\phi_1, \phi_2, ..., \phi_N)^T$ is the sought vector of the solution over the grid, $\mathbf{f} = (f_1, f_2, ..., f_N)^T$ is the right hand side vector, and $N \times N$ matrix

$$\mathbf{A} = \begin{pmatrix} b_1 & c_1 & \mu & 0 & 0 & 0 & 0\text{....................}0 \\ a_2 & b_2 & c_2 & 0 & 0 & 0 & 0\text{....................}0 \\ 0 & a_3 & b_3 & c_3 & 0 & 0 & 0\text{....................}0 \\ 0 & 0 & a_4 & b_4 & c_4 & 0 & 0\text{....................}0 \\ \dots & & & & & & \\ 0 & 0 & \text{...............}0 & a_{N-2} & b_{N-2} & c_{N-2} & 0 \\ 0 & 0 & \text{...............}0 & 0 & a_{N-1} & b_{N-1} & c_{N-1} \\ 0 & 0 & \text{...............}0 & 0 & \eta & a_N & b_N \end{pmatrix} \tag{64}$$

appears to be almost tridiagonal except for the two extraneous elements $A_{1,3} = \mu$ and $A_{N,N-2} = \eta$. The main diagonal elements are

$$b_1 = \frac{p_1}{h^2} - \frac{q_1}{h} + r_1, \quad b_k = r_k - \frac{2p_k}{h^2}, \; k \in \overline{2,(N-1)}, \text{ and } b_N = \frac{p_N}{h^2} + \frac{q_N}{h} + r_N, \tag{65}$$

the lower diagonal elements are

$$a_k = \frac{p_k}{h^2} - \frac{q_k}{2h}, \; k \in \overline{2,(N-1)}, \text{ and } a_N = -\frac{2p_N}{h^2} - \frac{q_N}{h}, \tag{66}$$

the upper diagonal elements are

$$c_1 = \frac{q_1}{h} - \frac{2p_1}{h^2}, \text{ and } c_k = \frac{p_k}{h^2} + \frac{q_k}{2h}, \; k \in \overline{2,(N-1)}, \tag{67}$$

and the remaining two elements are

$$\mu = \frac{p_1}{h^2} \text{ and } \eta = \frac{p_N}{h^2}. \tag{68}$$

To make the system truly tridiagonal, we premultiply both sides of (63) by $N \times N$ almost unity matrix

$$\mathbf{Q} = \begin{pmatrix} 1 & (-\mu/c_2) & 0 & 0 & 0\text{..................................}0 \\ 0 & 1 & 0 & 0 & 0\text{..................................}0 \\ 0 & 0 & 1 & 0 & 0\text{..................................}0 \\ \dots & & & & \\ 0 & 0 & \text{..................................}1 & 0 & 0 \\ 0 & 0 & \text{..................................}0 & 1 & 0 \\ 0 & 0 & \text{..................................}0 & (-\eta/a_{N-1}) & 1 \end{pmatrix}. \tag{69}$$

The resulting system readily soluble by double sweep method is

$$\tilde{\mathbf{A}}\boldsymbol{\phi} = \tilde{\mathbf{f}}, \tag{70}$$

where

$$\tilde{\mathbf{A}} = \mathbf{QA} = \begin{pmatrix} \tilde{b}_1 & \tilde{c}_1 & 0 & 0 & 0 & 0 & 0 \ldots\ldots\ldots\ldots 0 \\ a_2 & b_2 & c_2 & 0 & 0 & 0 & 0 \ldots\ldots\ldots\ldots 0 \\ 0 & a_3 & b_3 & c_3 & 0 & 0 & 0 \ldots\ldots\ldots\ldots 0 \\ 0 & 0 & a_4 & b_4 & c_4 & 0 & 0 \ldots\ldots\ldots\ldots 0 \\ \ldots\ldots\ldots\ldots\ldots\ldots\ldots\ldots\ldots\ldots\ldots\ldots\ldots\ldots\ldots \\ 0 & 0 & \ldots\ldots 0 & a_{N-2} & b_{N-2} & c_{N-2} & 0 \\ 0 & 0 & \ldots\ldots 0 & 0 & a_{N-1} & b_{N-1} & c_{N-1} \\ 0 & 0 & \ldots\ldots 0 & 0 & 0 & \tilde{a}_N & \tilde{b}_N \end{pmatrix}, \tag{71}$$

$$\tilde{b}_1 = b_1 - \frac{\mu}{c_2} a_2, \quad \tilde{c}_1 = c_1 - \frac{\mu}{c_2} b_2, \quad \tilde{a}_N = a_N - \frac{\eta}{a_{N-1}} b_{N-1}, \quad \tilde{b}_N = b_N - \frac{\eta}{a_{N-1}} c_{N-1}, \tag{72}$$

and

$$\tilde{\mathbf{f}} = \mathbf{Qf} = \left( \left( f_1 - \frac{\mu}{c_2} f_2 \right), f_2, f_3, \ldots, f_{N-1}, \left( f_N - \frac{\eta}{a_{N-1}} f_{N-1} \right) \right)^T. \tag{73}$$

It is possible to introduce the boundary condition analogous to $\phi(u) \xrightarrow[u\to\infty]{} 0$ so that the system remains tridiagonal. If $u_R \gg u_0$, a reasonable approximation of this boundary condition is $\phi_N = \phi(u_R) = 0$. This corresponds to $\eta = 0$, $a_N = 0$, $b_N = 1$, and $f_N = 0$. There is no specific boundary condition at $u = u_L$. This means that the numerical solution of (70) will represent a particular solution of (39), analogous to the analytical result (54) with $K_2 = 0$ and an indefinite value of $K_1$ or $\kappa$, which, in principle, can be determined using the normalization condition. However, at high velocities, roughly $u \in [u_0/2,\ 2u_0]$, where solution (44) is small, the particular solution analogous to (52) will dominate, which is determined by the source function parameters. Thus, if we are interested only in the high energy tail of the distribution, but not in the low energy part, there is no need to look for a definite value of $K_1$ or $\kappa$.

The numerical solution should coincide with the analytical result obtained for the case of isothermal target plasma in the previous subsection. Fig. 1 shows calculation results for the parameters given in Table I and Table II as an example. We show the energy distribution function $\frac{4\pi}{m_\alpha}\sqrt{\frac{2E}{m_\alpha}}\phi(u)$ versus $E = \frac{m_\alpha v_c^2 u^2}{2}$ instead of the solution $\phi(u)$ itself, since it is more apprehensible from the practical viewpoint. The exact analytical solution shown by a solid gray curve corresponds to formulas (52)-(54) with $K_2 = 0$ and $K_1$ given by (57), where the dimensionless parameter $\kappa = 2.4$. The normalization condition is expressed by (55). Note that, as it can be seen from (25), at $u = 0$ function $\phi(u) \propto -u^{-1}$ goes to minus infinity because of the second term in (52). This singularity formally takes place due to the use of spherical polar coordinates in velocity space. The probability density for velocity magnitude $\propto u^2\phi(u)$ and the probability density for kinetic energy $\propto u\phi(u)$ are both finite, since they include the appropriate Jacobian.

The numerical solution of the tridiagonal system of linear algebraic equations (70) shown by the dashed curve in Fig. 1 was obtained for $u_L = 8\times10^{-2}$, $u_R = 8$, and $N = 4096$. The solid black curve shows the simplified solution (38) with $b(u)$ given by (27). Thus, the exact analytical solution for the isothermal target plasma and the corresponding numerical solution coincide at high energies. Their Maxwellian parts at low energies also coincide for the chosen value of parameter $\kappa$. The slowing-down solution (38) fails to describe the high energy tail of the distribution correctly and is intrinsically inapplicable to demonstrate the Maxwellization process, since important physical properties are missing in the simplified equation neglecting the diffusion tensor.

Table I. Test particle source parameters.

| | |
|---|---|
| Species | Deuterons |
| Charge number | $Z_\alpha = 1$ |
| Mass | $m_\alpha = 3.344\times10^{-24}$ g |
| Injection energy | $E_0 = 150$ keV |
| Source rate | $S_0 = 0.1\times10^{14}$ cm$^{-3}$s$^{-1}$ |
| Width parameter | $\Delta = 10^{-3}$ |

Table II. Target plasma species.

| | |
|---|---|
| Electrons | $Z_e = -1$, $m_e = 9.109\times10^{-28}$ g<br>$n_e = 1.8\times10^{14}$ cm$^{-3}$, $T_e = 5$ keV |
| Helions | $Z_{He_2^3} = 2$, $m_{He_2^3} = 5.006\times10^{-24}$ g<br>$n_{He_2^3} = 0.6\times10^{14}$ cm$^{-3}$, $T_{He_2^3} = 5$ keV |
| Alphas | $Z_{He_2^4} = 2$, $m_{He_2^4} = 6.645\times10^{-24}$ g<br>$n_{He_2^4} = 0.2\times10^{14}$ cm$^{-3}$, $T_{He_2^4} = 5$ keV |
| Protons | $Z_p = 1$, $m_p = 1.673\times10^{-24}$ g<br>$n_p = 0.2\times10^{14}$ cm$^{-3}$, $T_p = 5$ keV |

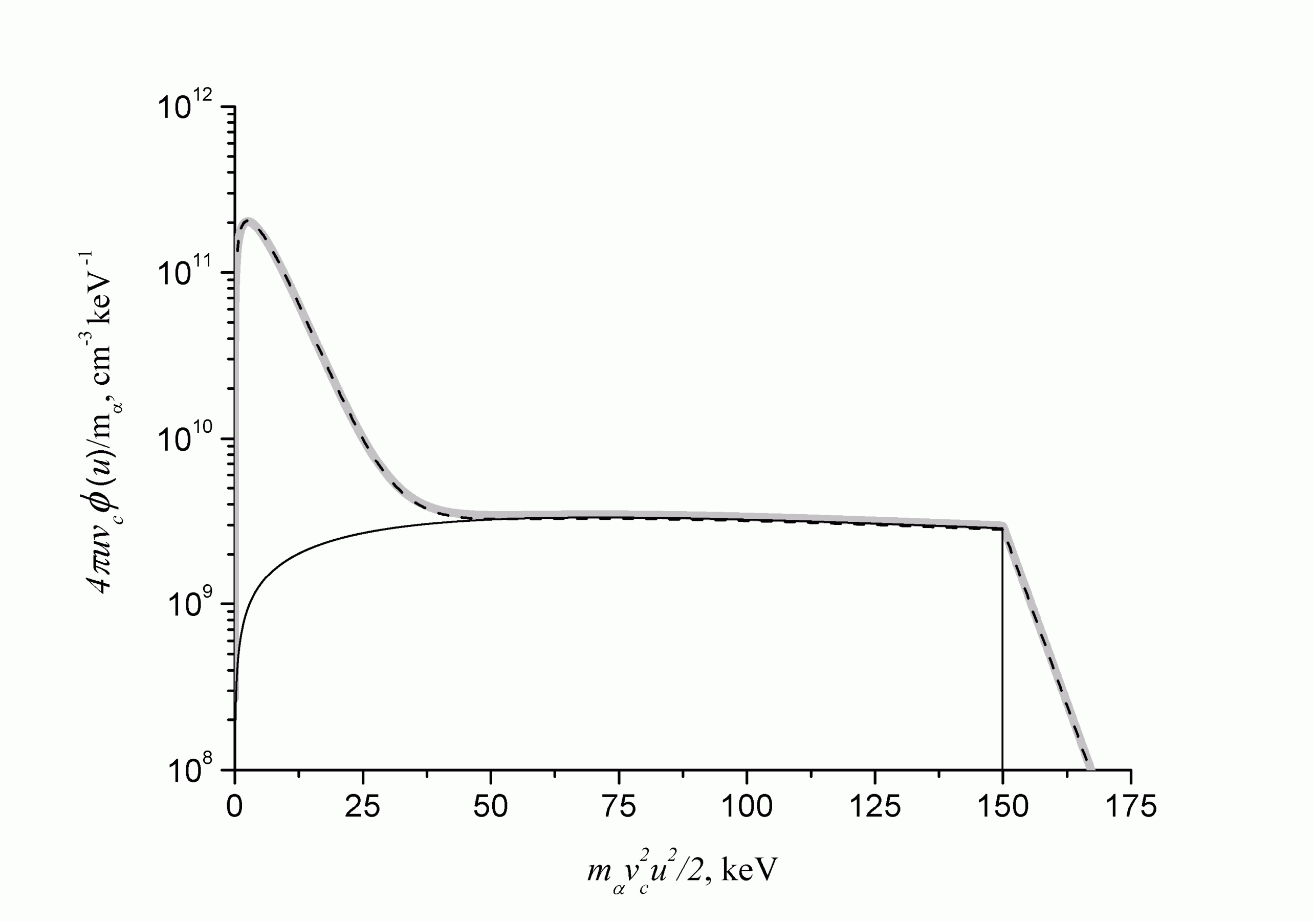

Fig. 1. Exact analytical solution (solid gray curve), numerical solution (dashed black curve) and simplified analytical solution (solid black curve) of the isotropic problem for 150 keV deuterons injected into isothermal $He_2^3 : He_2^4 : H_1^1$ (0.6 : 0.2 : 0.2) plasma at $T = 5$ keV.

## 5. Anisotropic problem

### 5.1. Slowing-down and pitch angle scattering

In this subsection we obtain an analytical steady state solution to equation (34) with $a(u) = 0$ and anisotropic source function (36). Differential operator

$$\mathcal{L} = \frac{\partial}{\partial \zeta}\left(1 - \zeta^2\right)\frac{\partial}{\partial \zeta} \tag{74}$$

is the one that occurs in Legendre equation. Its independent solutions are Legendre functions of the first kind $P_n(\zeta)$ and Legendre functions of the second kind $Q_n(\zeta)$. Since the latter are singular at $\zeta = \pm 1$ (i.e. $\vartheta = 0$ and $\vartheta = \pi$), it is meaningful to search for a solution in the form of an expansion

$$\phi(u,\zeta)=\sum_{n=0}^{\infty}\phi_n(u)P_n(\zeta) \tag{75}$$

suggested in [11] and applied in [6,7]. In contrast to [6,7], we do not hasten to simplify the equation, and obtain the solution in general form suitable for $b(u)$ and $c(u)$ given either by exact formulas (15), (16), (20), and (21), or by simplified formulas (27), (28). Substituting (75) into (34) with $a(u)=0$ and using the identity

$$\mathcal{L}P_n(\zeta)=-n(n+1)P_n(\zeta) \tag{76}$$

leads to equation

$$\sum_{n=0}^{\infty}\left(b(u)\frac{\partial\phi_n}{\partial u}+\frac{\partial b}{\partial u}\phi_n(u)-n(n+1)c(u)\phi_n(u)\right)P_n(\zeta)=-\frac{S_0\tau_s}{2\pi v_c^3}\delta(u-u_0)\mathcal{Z}(\zeta)\,. \tag{77}$$

Multiplying both sides of (77) by $P_m(\zeta)$, integrating over $[-1,1]$, using the orthogonality condition

$$\int_{-1}^{1}P_n(\zeta)P_m(\zeta)d\zeta=\frac{2}{2n+1}\delta_{mn}\,, \tag{78}$$

where $\delta_{mn}$ is Kronecker symbol, and denoting

$$\mathcal{Z}_n=\int_{-1}^{1}\mathcal{Z}(\zeta)P_n(\zeta)d\zeta \tag{79}$$

we arrive at first order ordinary differential equation

$$\frac{\partial\phi_n}{\partial u}+\left(\frac{1}{b(u)}\frac{\partial b}{\partial u}-\frac{n(n+1)c(u)}{b(u)}\right)\phi_n(u)=-\frac{2n+1}{2}\frac{S_0\tau_s}{2\pi v_c^3}\frac{\mathcal{Z}_n}{b(u)}\delta(u-u_0)\,. \tag{80}$$

The general solution of the corresponding homogeneous equation obtained by variable separation method is

$$\phi_n(u)=\frac{A}{b(u)}\exp\left(\int_{u_0}^{u}\frac{n(n+1)c(\tilde{u})}{b(\tilde{u})}d\tilde{u}\right), \tag{81}$$

where $A$ is an arbitrary constant. The solution of inhomogeneous equation (80) can be obtained by variation of constant $A$. Regarding it as an unknown function $A(u)$, and substituting (81) into (80) yields the derivative

$$A'(u)=-\frac{2n+1}{2}\frac{S_0\tau_s}{2\pi v_c^3}\mathcal{Z}_n\delta(u-u_0)\exp\left(-\int_{u_0}^{u}\frac{n(n+1)c(\tilde{u})}{b(\tilde{u})}d\tilde{u}\right). \tag{82}$$

Thus, $A'(u)=0$ everywhere except $u=u_0$, where the exponent in (82) equals unity. Therefore, integrating (82), we have

$$A(u)=\frac{2n+1}{2}\frac{S_0\tau_s}{2\pi v_c^3}\mathcal{Z}_n H(u_0-u)+\tilde{A}, \tag{83}$$

where $\tilde{A}$ is an arbitrary constant. Assuming $\phi_n(u)\xrightarrow[u\to\infty]{}0$, we find $\tilde{A}=0$. Finally, the solution of (80) is

$$\phi_n(u)=\frac{S_0\tau_s}{4\pi v_c^3}(2n+1)\mathcal{Z}_n\frac{H(u_0-u)}{b(u)}\exp\left(-n(n+1)\int_u^{u_0}\frac{c(\tilde{u})}{b(\tilde{u})}d\tilde{u}\right). \tag{84}$$

Note that $\phi_0(u)$ coincides with (38) because $\mathcal{Z}(\zeta)$ is normalized to unity, and $P_0(\zeta)=1$. If the source is monodirectional, and the injection angle cosine is $\zeta_0=\cos\vartheta_0$, i.e.

$$\mathcal{Z}(\zeta)=\delta(\zeta-\zeta_0), \tag{85}$$

then (79) gives

$$\mathcal{Z}_n=P_n(\zeta_0). \tag{86}$$

### 5.2. Complete equation with slowing-down, velocity diffusion, and pitch angle scattering

A semi-analytical steady state solution to equation (34), including $a(u)$, with anisotropic source function (36) can be obtained in the form (75). Applying the procedure similar to (76)-(79), we arrive at second order ordinary differential equation

$$\frac{a(u)}{2u^3}\frac{\partial^2\phi_n}{\partial u^2}+\left(\frac{b(u)}{u^2}-\frac{a(u)}{2u^4}+\frac{1}{2u^3}\frac{\partial a}{\partial u}\right)\frac{\partial\phi_n}{\partial u}+\left(\frac{1}{u^2}\frac{\partial b}{\partial u}-n(n+1)\frac{c(u)}{u^2}\right)\phi_n(u)$$
$$=-\frac{2n+1}{2}\frac{S_0\tau_s}{2\pi v_c^3}\frac{\mathcal{Z}_n}{u^2}\delta(u-u_0). \tag{87}$$

To solve it numerically, we replace $\delta(u-u_0)$ with $\mathcal{D}(u-u_0)$ given by (58), and rewrite (87) as

$$p(u)\frac{\partial^2\phi_n}{\partial u^2}+q(u)\frac{\partial\phi_n}{\partial u}+r(u)\phi_n(u)=f(u), \tag{88}$$

where

$$p(u)=\frac{a(u)}{2u^3}, \tag{89}$$

$$q(u)=\frac{b(u)}{u^2}-\frac{a(u)}{2u^4}+\frac{1}{2u^3}\frac{\partial a}{\partial u}, \tag{90}$$

$$r(u)=\frac{1}{u^2}\frac{\partial b}{\partial u}-n(n+1)\frac{c(u)}{u^2}, \tag{91}$$

$$f(u) = -\frac{S_0 \tau_s}{4\pi v_c^3} \frac{(2n+1)\mathcal{Z}_n}{u^2} \frac{1}{\Delta\sqrt{\pi}} e^{-(u-u_0)^2/\Delta^2} . \tag{92}$$

Equation (88) is formally analogous to (39), thus, we can apply the numerical method described in subsection 4.3 to obtain a solution $\phi_n(u)$ over a uniform grid on a finite interval $[u_L, u_R]$. After that the final result is calculated using (75). Each term in the series requires equation (88) to be solved numerically. The summation of converging series is performed until the required relative precision is achieved. A successful verification of the algorithm was performed as described below.

Note that for $n = 0$ the problem expressed by (88)-(92) reduces to (39)-(43), and the exact analytical solution obtained in subsection 4.2 is valid. It can be used to verify the numerical algorithm. Another possible way of verification is to artificially reduce $a(u)$ multiplying it by a small constant, e.g. $10^{-2}$, and obtain a complete semi-analytical solution in this special case. The result should agree with the simplified analytical solution of subsection 5.1 obtained for $a(u) = 0$.

Fig. 2 shows calculation results for the parameters given in Table I and Table II as an example. The source is monodirectional, so that (85) and (86) hold. Injection angle is $\vartheta_0 = 0°$ in this example. We show the energy distribution function $\frac{1}{m_\alpha}\sqrt{\frac{2E}{m_\alpha}}\phi(u,\zeta)$ versus $E = \frac{m_\alpha v_c^2 u^2}{2}$ instead of the solution $\phi(u,\zeta)$ itself. Numerical solutions of (88) were obtained for $u_L = 4\times10^{-2}$, $u_R = 8$, and grid dimension $N = 4096$. Function $\phi(u,\zeta)$ was calculated by (75). Solid gray curves show the complete semi-analytical solution corresponding to (75) and (88), taking into account slowing-down, velocity diffusion, and pitch angle scattering. Dashed black curves show the simplified analytical solution corresponding to (75) and (84), taking into account slowing-down and pitch angle scattering, and using (27), (28) to calculate $b(u)$ and $c(u)$. The simplified solution at all angles fails to describe high energy tails of the distribution. Besides, it is essentially unable to demonstrate the Maxwellization process observed in the low energy part of the correct distribution.

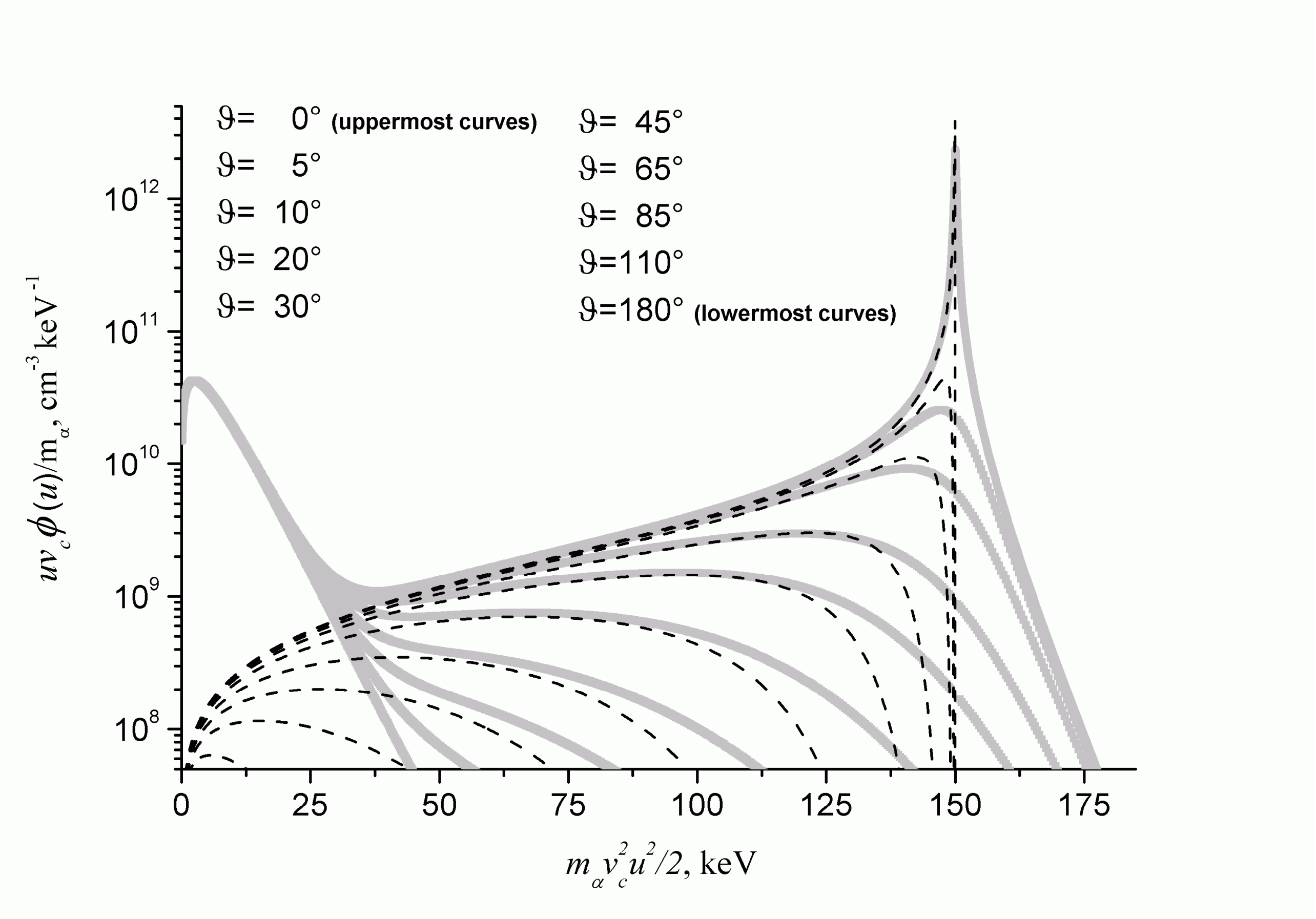

Fig. 2. Complete semi-analytical solution (solid gray curves), and simplified analytical solution (dashed black curves) of the anisotropic problem for 150 keV deuterons injected into isothermal $He_2^3 : He_2^4 : H_1^1$ (0.6 : 0.2 : 0.2) plasma at $T = 5$ keV.

# 6. Time-dependent problem

## 6.1. Slowing-down and pitch angle scattering

In this subsection the nonstationary problem is solved. Consider equation (34) with time-dependent source function

$$S_\alpha(u,\zeta,\tau) = \frac{S_0}{2\pi v_c^3}\frac{1}{u^2}\delta(u-u_0)\mathcal{Z}(\zeta)\big(H(\tau-\tau_0)-H(\tau-\tau_1)\big), \tag{93}$$

where $\tau_0$ and $\tau_1$ designate the source action start and stop times respectively, and $0<\tau_0<\tau_1$.

A simplified equation with $a(u)=0$ can be readily solved using analytical techniques. As opposed to [7], we obtain the solution in general form suitable for $b(u)$ and $c(u)$ given either by exact formulas (15), (16), (20), and (21), or by simplified formulas (27), (28). Assuming the initial condition

$$\phi(u,\zeta,\tau)\big|_{\tau=0}=0\,, \tag{94}$$

expanding

$$\phi(u,\zeta,\tau)=\sum_{n=0}^{\infty}\phi_n(u,\tau)P_n(\zeta)\,, \tag{95}$$

recalling (76) and (78), and applying Laplace transform

$$\phi_n(u,p)=\int_0^{\infty}e^{-p\tau}\phi_n(u,\tau)d\tau\,, \tag{96}$$

we reduce (34) with $a(u)=0$ to first order ordinary differential equation

$$\frac{\partial\phi_n}{\partial u}+\left(\frac{1}{b(u)}\frac{\partial b}{\partial u}-\frac{pu^2}{b(u)}-\frac{n(n+1)c(u)}{b(u)}\right)\phi_n(u,p)=-\frac{S_0\tau_s}{4\pi v_c^3}\frac{(2n+1)\mathcal{Z}_n\delta(u-u_0)}{b(u)}\frac{\left(e^{-p\tau_0}-e^{-p\tau_1}\right)}{p}. \tag{97}$$

Next, we solve the corresponding homogeneous equation by variable separation method and then by variation of constant we find

$$\phi_n(u,p)=\frac{S_0\tau_s}{4\pi v_c^3}\frac{(2n+1)\mathcal{Z}_nH(u_0-u)}{b(u)}e^{-\int_u^{u_0}\frac{n(n+1)c(\tilde{u})}{b(\tilde{u})}d\tilde{u}}\frac{1}{p}\left(e^{-p\left(\tau_0+\int_u^{u_0}\frac{\tilde{u}^2d\tilde{u}}{b(\tilde{u})}\right)}-e^{-p\left(\tau_1+\int_u^{u_0}\frac{\tilde{u}^2d\tilde{u}}{b(\tilde{u})}\right)}\right). \tag{98}$$

Since $u<u_0$ region is considered in this simplified problem with no velocity diffusion, and $u^2/b(u)>0$, the integral $\int_u^{u_0}\frac{\tilde{u}^2d\tilde{u}}{b(\tilde{u})}>0$, and thus we can use the known Laplace transform

$$\frac{1}{p}e^{-\alpha p}\to H(\tau-\alpha)\ \text{for}\ \alpha>0\,. \tag{99}$$

Finally, the time-dependent solution is

$$\phi_n(u,\tau)=\frac{S_0\tau_s}{4\pi v_c^3}\frac{(2n+1)\mathcal{Z}_nH(u_0-u)e^{-n(n+1)\int_u^{u_0}\frac{c(\tilde{u})}{b(\tilde{u})}d\tilde{u}}}{b(u)}\left(H\left(\tau-\tau_0-\int_u^{u_0}\frac{\tilde{u}^2d\tilde{u}}{b(\tilde{u})}\right)-H\left(\tau-\tau_1-\int_u^{u_0}\frac{\tilde{u}^2d\tilde{u}}{b(\tilde{u})}\right)\right). \tag{100}$$

For $n=0$, $\tau_0=0$, and $\tau_1=\infty$ function

$$\phi_0(u,\tau)=\frac{S_0\tau_s}{4\pi v_c^3}\frac{H(u_0-u)}{b(u)}H\left(\tau-\int_u^{u_0}\frac{\tilde{u}^2d\tilde{u}}{b(\tilde{u})}\right) \tag{101}$$

is the nonstationary slowing-down solution of (34) with $a(u)=0$ and $c(u)=0$, which is similar to nonstationary solutions obtained in [8,9].

Since the time-dependent Heaviside step function equals unity for $\tau \to \infty$, this passage to the limit makes (100) coincide with steady state solution (84), and also makes (101) coincide with the simplest steady state slowing-down distribution (38).

### 6.2. Complete equation with slowing-down, velocity diffusion, and pitch angle scattering

The problem (34), (93) is solved semi-analytically over a uniform grid $\tau_j = (j-1)h_\tau$, $j \in \overline{1,M}$, $h_\tau = T/(M-1)$, on a finite time interval $\tau \in [0,T]$ and a uniform grid $u_i = u_L + (i-1)h_u$, $i \in \overline{1,N}$, $h_u = (u_R - u_L)/(N-1)$, on a finite velocity interval $u \in [u_L, u_R]$, employing expansion (95) and applying Crank–Nicolson method, proposed in [12], to the equation for $\phi_n(u,\tau)$

$$\frac{\partial \phi_n}{\partial \tau} = \mathcal{F}\left(\phi_n, \frac{\partial \phi_n}{\partial u}, \frac{\partial^2 \phi_n}{\partial u^2} u, \tau\right), \tag{102}$$

where

$$\mathcal{F}\left(\phi_n, \frac{\partial \phi_n}{\partial u}, \frac{\partial^2 \phi_n}{\partial u^2} u, \tau\right) = p(u)\frac{\partial^2 \phi_n}{\partial u^2} + q(u)\frac{\partial \phi_n}{\partial u} + r(u)\phi_n(u) + f(u,\tau), \tag{103}$$

$$p(u) = \frac{a(u)}{2u^3}, \tag{104}$$

$$q(u) = \frac{b(u)}{u^2} - \frac{a(u)}{2u^4} + \frac{1}{2u^3}\frac{\partial a}{\partial u}, \tag{105}$$

$$r(u) = \frac{1}{u^2}\frac{\partial b}{\partial u} - n(n+1)\frac{c(u)}{u^2}, \tag{106}$$

$$f(u,\tau) = \frac{S_0 \tau_s}{4\pi v_c^3}\frac{(2n+1)\mathcal{Z}_n}{u^2}\frac{1}{\Delta\sqrt{\pi}} e^{-(u-u_0)^2/\Delta^2}\left(H(\tau-\tau_0) - H(\tau-\tau_1)\right). \tag{107}$$

The discrete counterpart of (102)

$$\frac{\phi_n^{i,j+1} - \phi_n^{i,j}}{h_\tau} = \frac{1}{2}\left(\mathcal{F}^{i,j+1}\left(\phi_n, \frac{\partial \phi_n}{\partial u}, \frac{\partial^2 \phi_n}{\partial u^2} u, \tau\right) - \mathcal{F}^{i,j}\left(\phi_n, \frac{\partial \phi_n}{\partial u}, \frac{\partial^2 \phi_n}{\partial u^2} u, \tau\right)\right) \tag{108}$$

involving forward difference derivatives at $u = u_1 = u_L$, i.e. for $i=1$, central difference derivatives at the inner velocity grid points, i.e. for $i \in \overline{2,(N-1)}$, and backward difference derivatives at $u = u_N = u_R$ , i.e. $i = N$ , after applying a simple tridiagonalization resembling (69), leads to a tridiagonal system of linear algebraic equations

$$\mathbf{A}\mathbf{\Phi}^{j+1} = -\left(\mathbf{B}\mathbf{\Phi}^{j} + \mathbf{D}^{j} + \mathbf{D}^{j+1}\right), \tag{109}$$

which needs to be solved at every step in time to get the solution vector at the subsequent time grid point $(j+1)$

$$\mathbf{\Phi}^{j+1} = \left(\phi_n^{1,j+1}, \phi_n^{2,j+1}, \ldots, \phi_n^{N,j+1}\right)^T, \tag{110}$$

using the solution vector $\mathbf{\Phi}^j$ previously obtained at the preceding time grid point $j$ and assuming zero initial condition.

Time-independent tridiagonal $N \times N$ matrices in (109) are

$$\mathbf{A} = \begin{pmatrix} \tilde{b}_1 & \tilde{c}_1 & 0 & 0 & 0 & 0 & 0 \ldots\ldots\ldots 0 \\ a_2 & \tilde{b}_2 & c_2 & 0 & 0 & 0 & 0 \ldots\ldots\ldots 0 \\ 0 & a_3 & \tilde{b}_3 & c_3 & 0 & 0 & 0 \ldots\ldots\ldots 0 \\ 0 & 0 & a_4 & \tilde{b}_4 & c_4 & 0 & 0 \ldots\ldots\ldots 0 \\ \ldots\ldots\ldots\ldots\ldots\ldots\ldots\ldots\ldots\ldots\ldots\ldots\ldots\ldots \\ 0 & 0 & \ldots\ldots 0 & a_{N-2} & \tilde{b}_{N-2} & c_{N-2} & 0 \\ 0 & 0 & \ldots\ldots 0 & 0 & a_{N-1} & \tilde{b}_{N-1} & c_{N-1} \\ 0 & 0 & \ldots\ldots 0 & 0 & 0 & \tilde{a}_N & \tilde{b}_N \end{pmatrix} \tag{111}$$

and

$$\mathbf{B} = \begin{pmatrix} \hat{b}_1 & \hat{c}_1 & 0 & 0 & 0 & 0 & 0 \ldots\ldots\ldots 0 \\ a_2 & \hat{b}_2 & c_2 & 0 & 0 & 0 & 0 \ldots\ldots\ldots 0 \\ 0 & a_3 & \hat{b}_3 & c_3 & 0 & 0 & 0 \ldots\ldots\ldots 0 \\ 0 & 0 & a_4 & \hat{b}_4 & c_4 & 0 & 0 \ldots\ldots\ldots 0 \\ \ldots\ldots\ldots\ldots\ldots\ldots\ldots\ldots\ldots\ldots\ldots\ldots\ldots\ldots \\ 0 & 0 & \ldots\ldots 0 & a_{N-2} & \hat{b}_{N-2} & c_{N-2} & 0 \\ 0 & 0 & \ldots\ldots 0 & 0 & a_{N-1} & \hat{b}_{N-1} & c_{N-1} \\ 0 & 0 & \ldots\ldots 0 & 0 & 0 & \hat{a}_N & \hat{b}_N \end{pmatrix}. \tag{112}$$

Main diagonal elements of these matrices are calculated as

$$\tilde{b}_i = h_\tau \left( \frac{r(u_i)}{2} - \frac{p(u_i)}{h_u^2} \right) - 1, \quad \hat{b}_i = h_\tau \left( \frac{r(u_i)}{2} - \frac{p(u_i)}{h_u^2} \right) + 1 \quad \text{for} \quad i \in \overline{2,(N-1)}. \tag{113}$$

Upper and lower diagonal elements are

$$a_i = \frac{h_\tau}{2}\left(\frac{p(u_i)}{h_u^2} - \frac{q(u_i)}{2h_u}\right), \quad c_i = \frac{h_\tau}{2}\left(\frac{p(u_i)}{h_u^2} + \frac{q(u_i)}{2h_u}\right) \quad \text{for} \quad i \in \overline{2,(N-1)}. \tag{114}$$

The remaining elements are given by

$$\tilde{b}_1 = \frac{h_\tau}{2}\left(\frac{p(u_1)}{h_u^2} - \frac{q(u_1)}{h_u} + r(u_1)\right) - 1 - \frac{h_\tau}{2}\frac{p(u_1)}{h_u^2}\frac{a_2}{c_2}, \tag{115}$$

$$\tilde{b}_N = \frac{h_\tau}{2}\left(\frac{p(u_N)}{h_u^2} + \frac{q(u_N)}{h_u} + r(u_N)\right) - 1 - \frac{h_\tau}{2}\frac{p(u_N)}{h_u^2}\frac{c_{N-1}}{a_{N-1}}, \tag{116}$$

$$\hat{b}_1 = \frac{h_\tau}{2}\left(\frac{p(u_1)}{h_u^2} - \frac{q(u_1)}{h_u} + r(u_1)\right) + 1 - \frac{h_\tau}{2}\frac{p(u_1)}{h_u^2}\frac{a_2}{c_2}, \tag{117}$$

$$\hat{b}_N = \frac{h_\tau}{2}\left(\frac{p(u_N)}{h_u^2} + \frac{q(u_N)}{h_u} + r(u_N)\right) + 1 - \frac{h_\tau}{2}\frac{p(u_N)}{h_u^2}\frac{c_{N-1}}{a_{N-1}}, \tag{118}$$

$$\tilde{a}_N = -h_\tau\left(\frac{p(u_N)}{h_u^2} + \frac{q(u_N)}{2h_u}\right) - \frac{h_\tau}{2}\frac{p(u_N)}{h_u^2}\frac{\tilde{b}_{N-1}}{a_{N-1}}, \quad \hat{a}_N = -h_\tau\left(\frac{p(u_N)}{h_u^2} + \frac{q(u_N)}{2h_u}\right) - \frac{h_\tau}{2}\frac{p(u_N)}{h_u^2}\frac{\hat{b}_{N-1}}{a_{N-1}}, \tag{119}$$

$$\tilde{c}_1 = h_\tau\left(\frac{q(u_1)}{2h_u} - \frac{p(u_1)}{h_u^2}\right) - \frac{h_\tau}{2}\frac{p(u_1)}{h_u^2}\frac{\tilde{b}_2}{c_2}, \quad \hat{c}_1 = h_\tau\left(\frac{q(u_1)}{2h_u} - \frac{p(u_1)}{h_u^2}\right) - \frac{h_\tau}{2}\frac{p(u_1)}{h_u^2}\frac{\hat{b}_2}{c_2}. \tag{120}$$

Time-dependent vector $\mathbf{D}^j = \left(D_1^j, D_2^j, ..., D_N^j\right)^T$ is calculated as

$$D_i^j = \frac{h_\tau}{2} f(u_i, \tau_j) \quad \text{for} \quad i \in \overline{2,(N-1)}, \tag{121}$$

$$D_1^j = \frac{h_\tau}{2}\left(f(u_1,\tau_j) - \frac{h_\tau}{2c_2}\frac{p(u_1)}{h_u^2} f(u_2,\tau_j)\right), \quad D_N^j = \frac{h_\tau}{2}\left(f(u_N,\tau_j) - \frac{h_\tau}{2a_{N-1}}\frac{p(u_N)}{h_u^2} f(u_{N-1},\tau_j)\right). \tag{122}$$

The resulting distribution function is calculated using (95). Each term in the series requires equation (109) to be solved numerically within the time loop. The summation of converging series is performed until the required relative precision is achieved. A successful quantitative verification of the algorithm was performed in two ways. Since the distribution function is normalized to test particle density, the integral of the distribution function over the entire velocity space should be equal to the source rate $S_0$ multiplied by the source operation duration $\tau_s(\tau - \tau_0)$, while the source (93) is acting (i.e. for $\tau_0 < \tau < \tau_1$). After the source termination (i.e. for $\tau > \tau_1$) the density of test particles should remain constant and equal to the source rate $S_0$ multiplied by the total operation duration $\tau_s(\tau_1 - \tau_0)$. Another way to verify the semi-analytical solution is to artificially reduce the function $a(u)$ responsible for velocity diffusion, multiplying it by a small constant, e.g. $10^{-2}$. In this special case the complete semi-analytical solution should agree with the exact analytical solution given by (95) and (100) obtained with $a(u) = 0$.

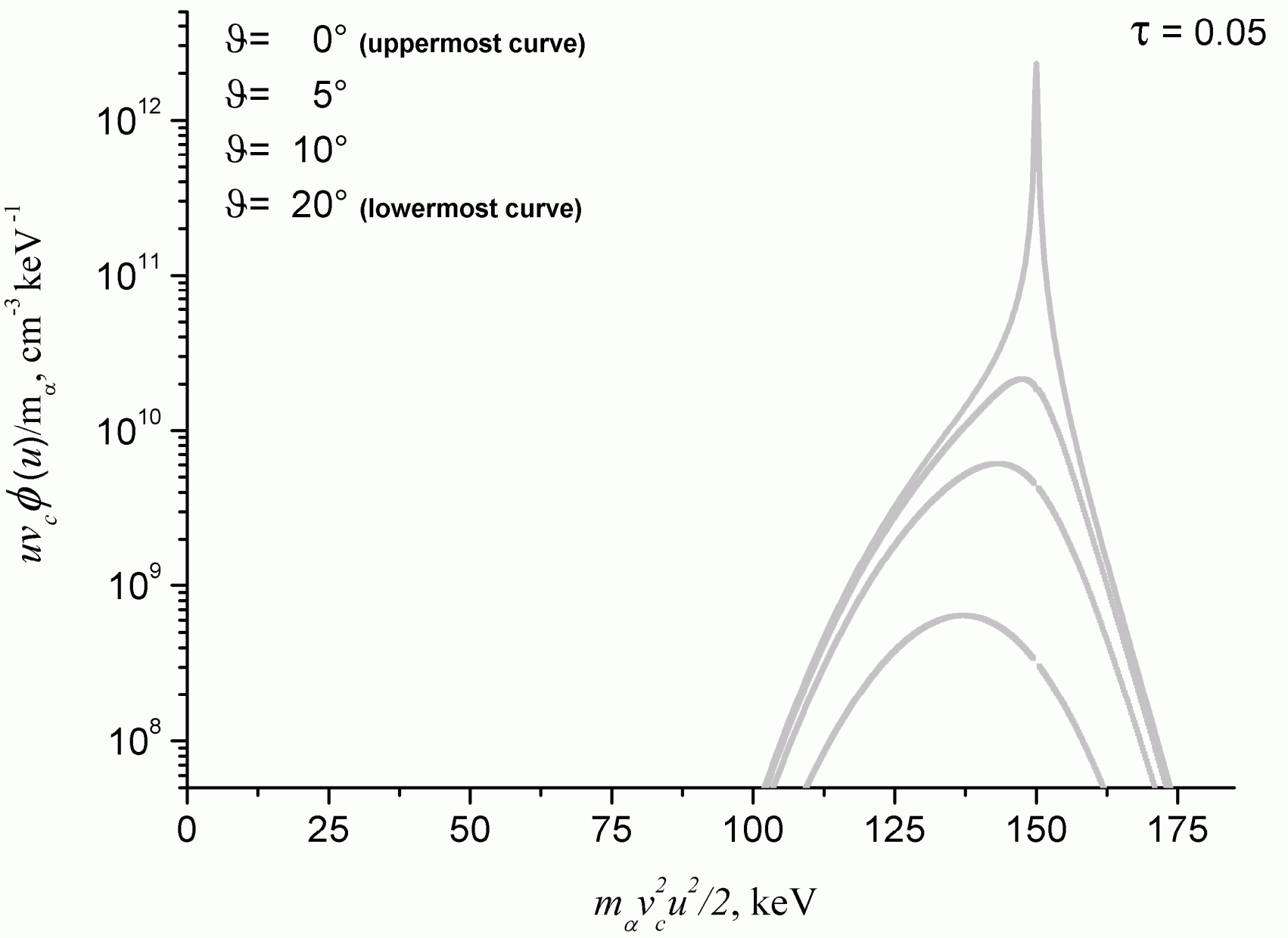
ϑ= 0° (uppermost curve)
ϑ= 5°
ϑ= 10°
ϑ= 20° (lowermost curve)
τ = 0.05
$uv_c\phi(u)/m_\alpha$, cm$^{-3}$keV$^{-1}$
$m_\alpha v_c^2 u^2/2$, keV

**(a)**

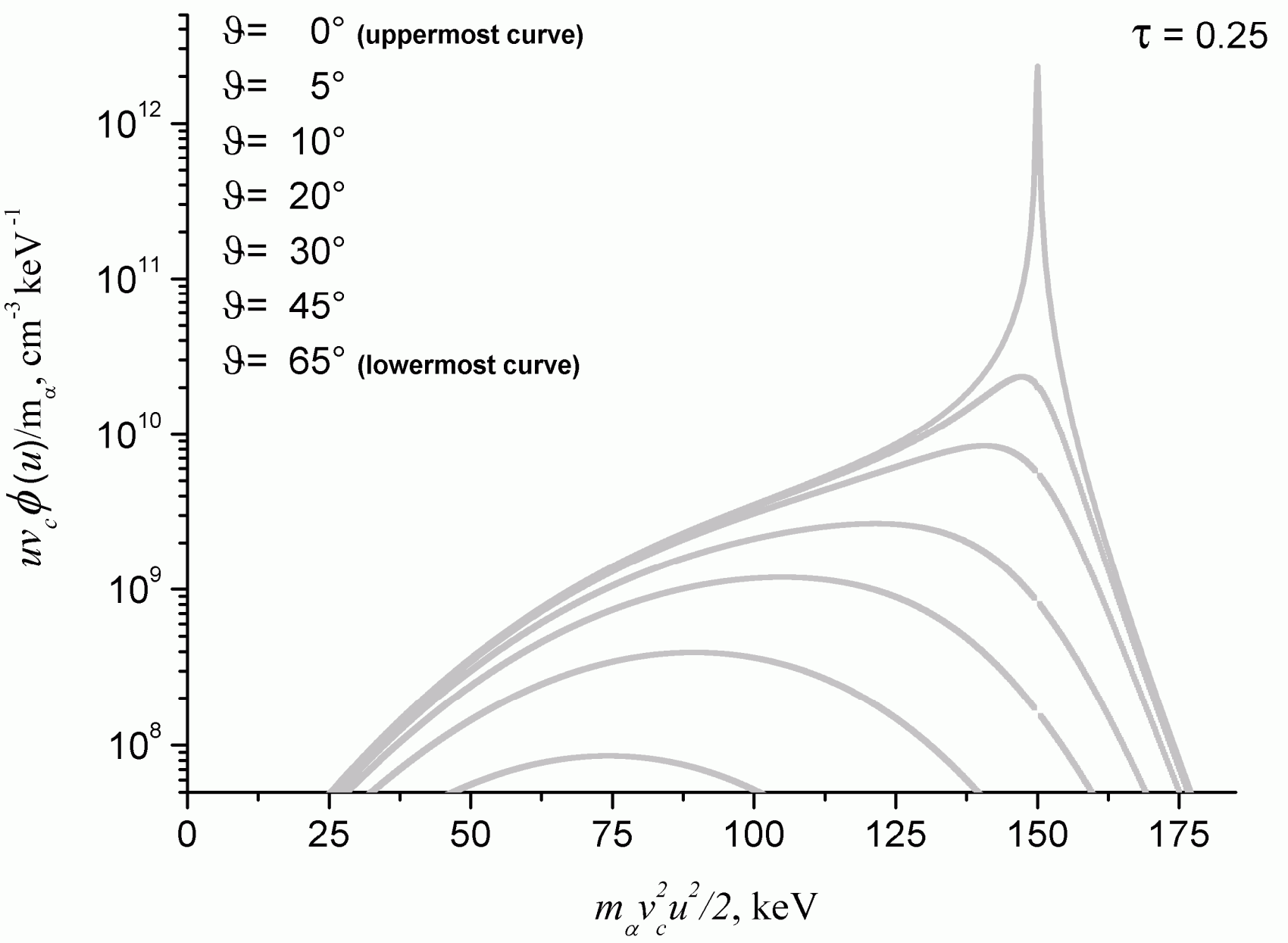
ϑ= 0° (uppermost curve)
ϑ= 5°
ϑ= 10°
ϑ= 20°
ϑ= 30°
ϑ= 45°
ϑ= 65° (lowermost curve)
τ = 0.25
$uv_c\phi(u)/m_\alpha$, cm$^{-3}$keV$^{-1}$
$m_\alpha v_c^2 u^2/2$, keV

**(b)**

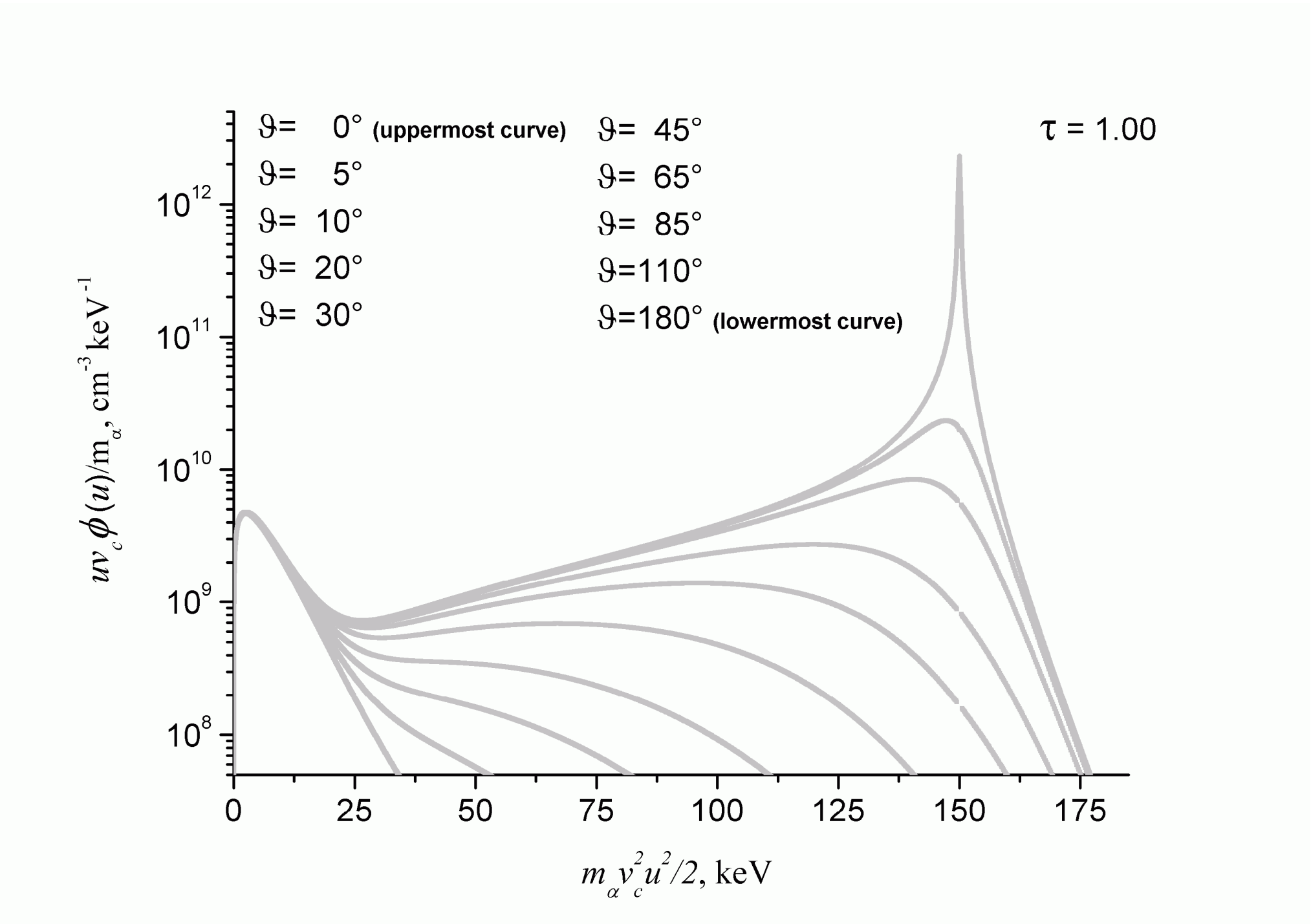

**(c)**

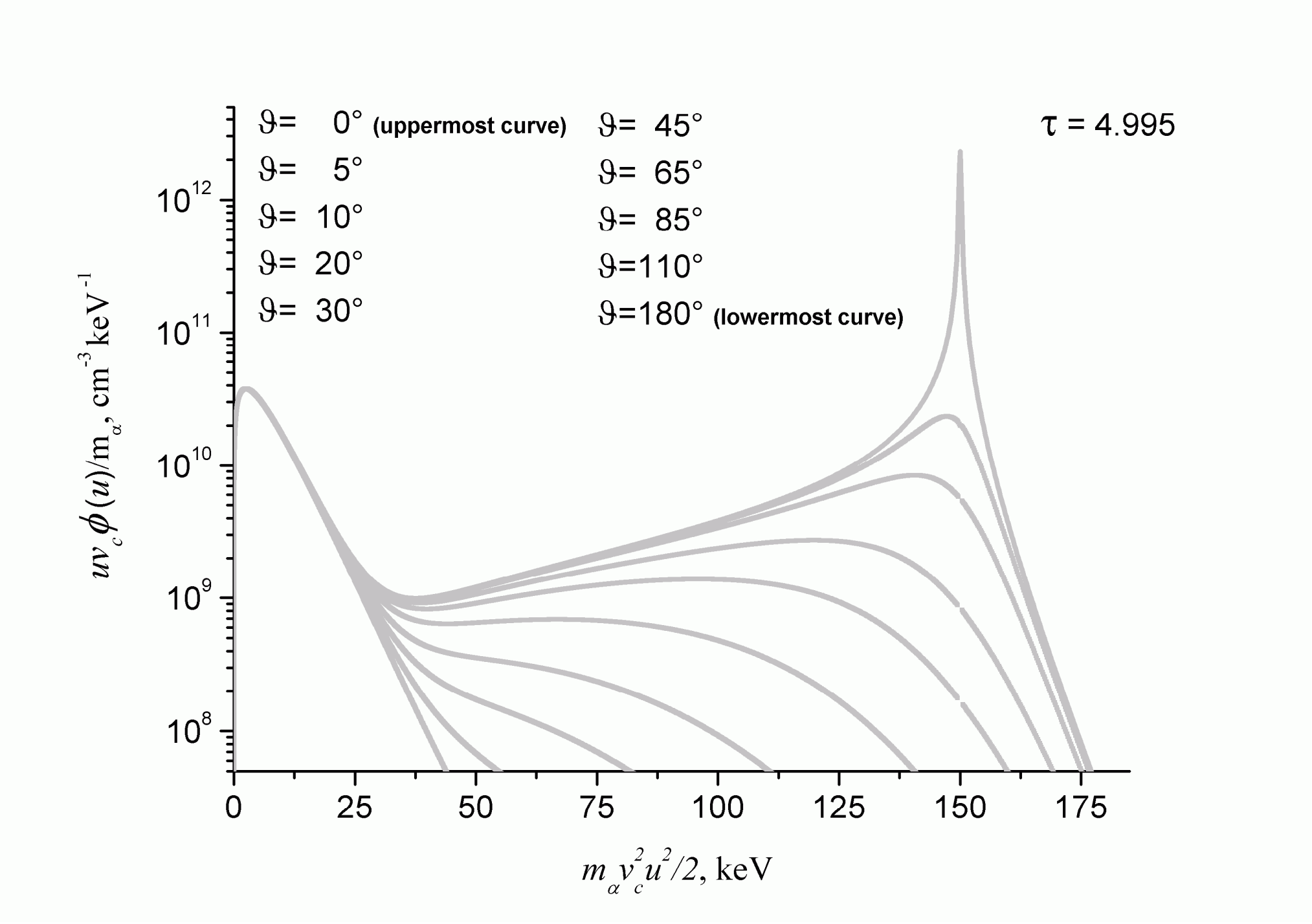

**(d)**

Fig. 3. Complete semi-analytical time-dependent solution of the anisotropic problem for 150 keV deuterons injected into isothermal $He_2^3 : He_2^4 : H_1^1$ (0.6 : 0.2 : 0.2) plasma at $T$ = 5 keV. Evolution during the source action (a) for $\tau = 0.05$, (b) for $\tau = 0.25$, (c) for $\tau = 1.00$, and (d) for $\tau = 4.995$.

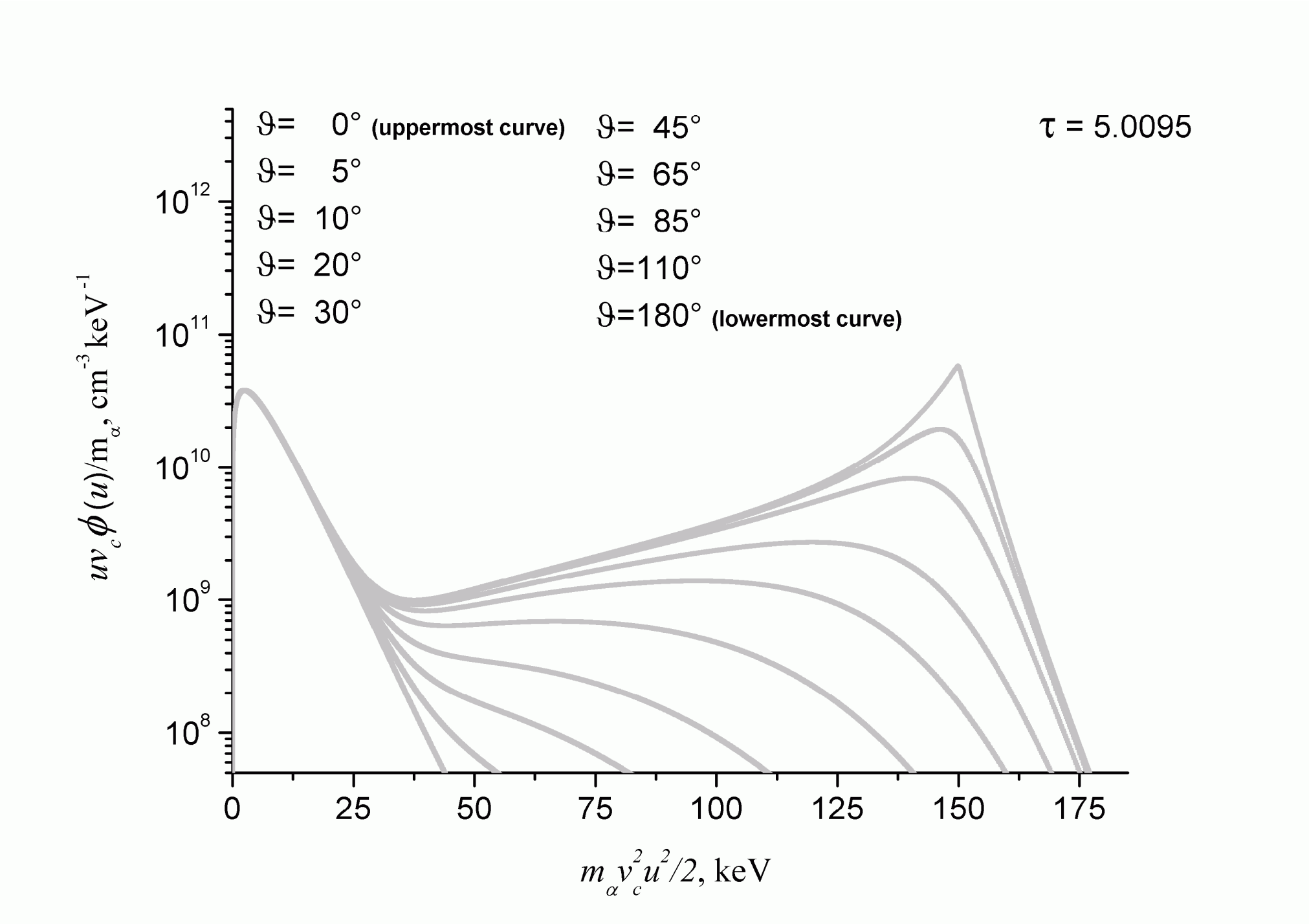

**(a)**

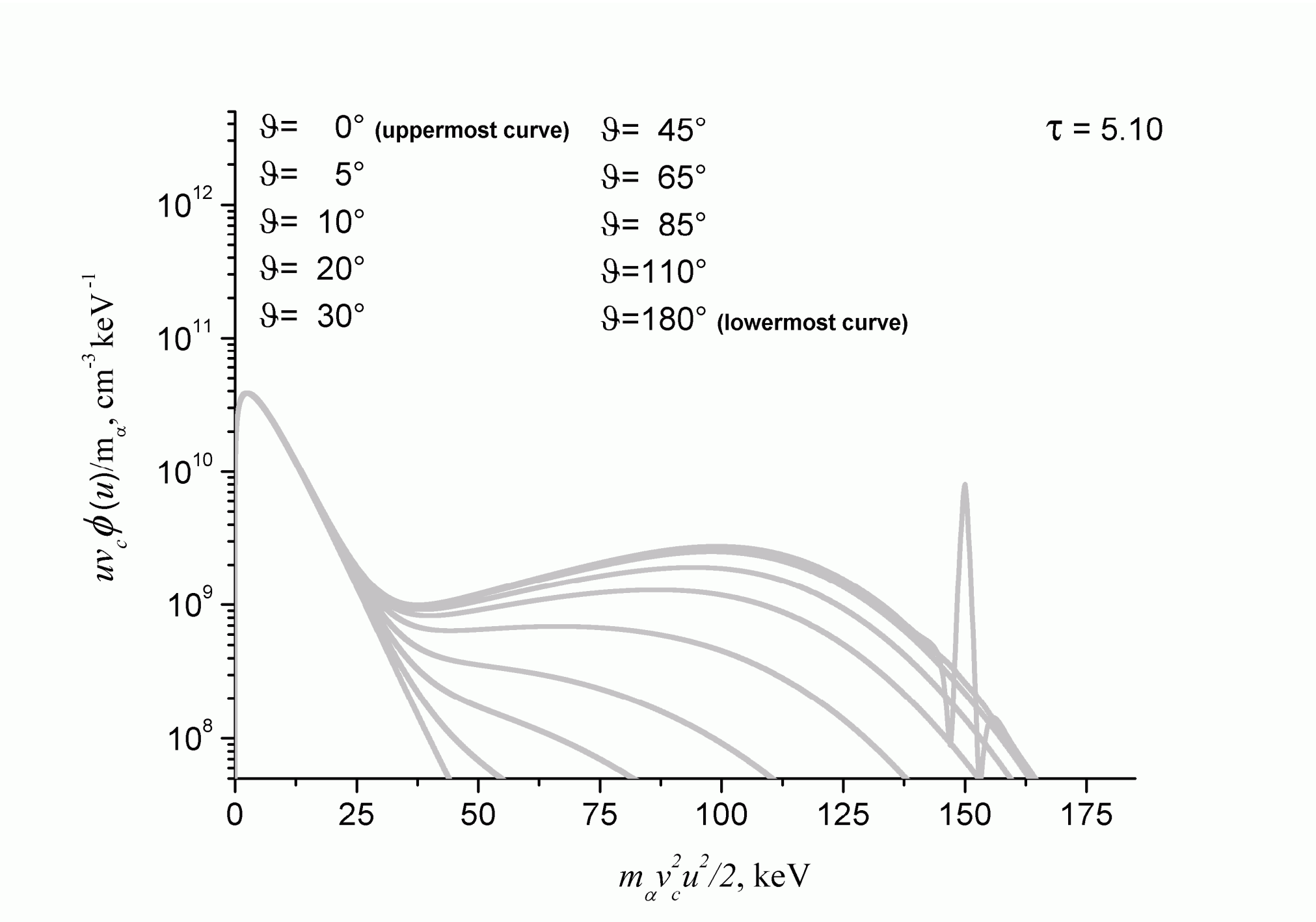

**(b)**

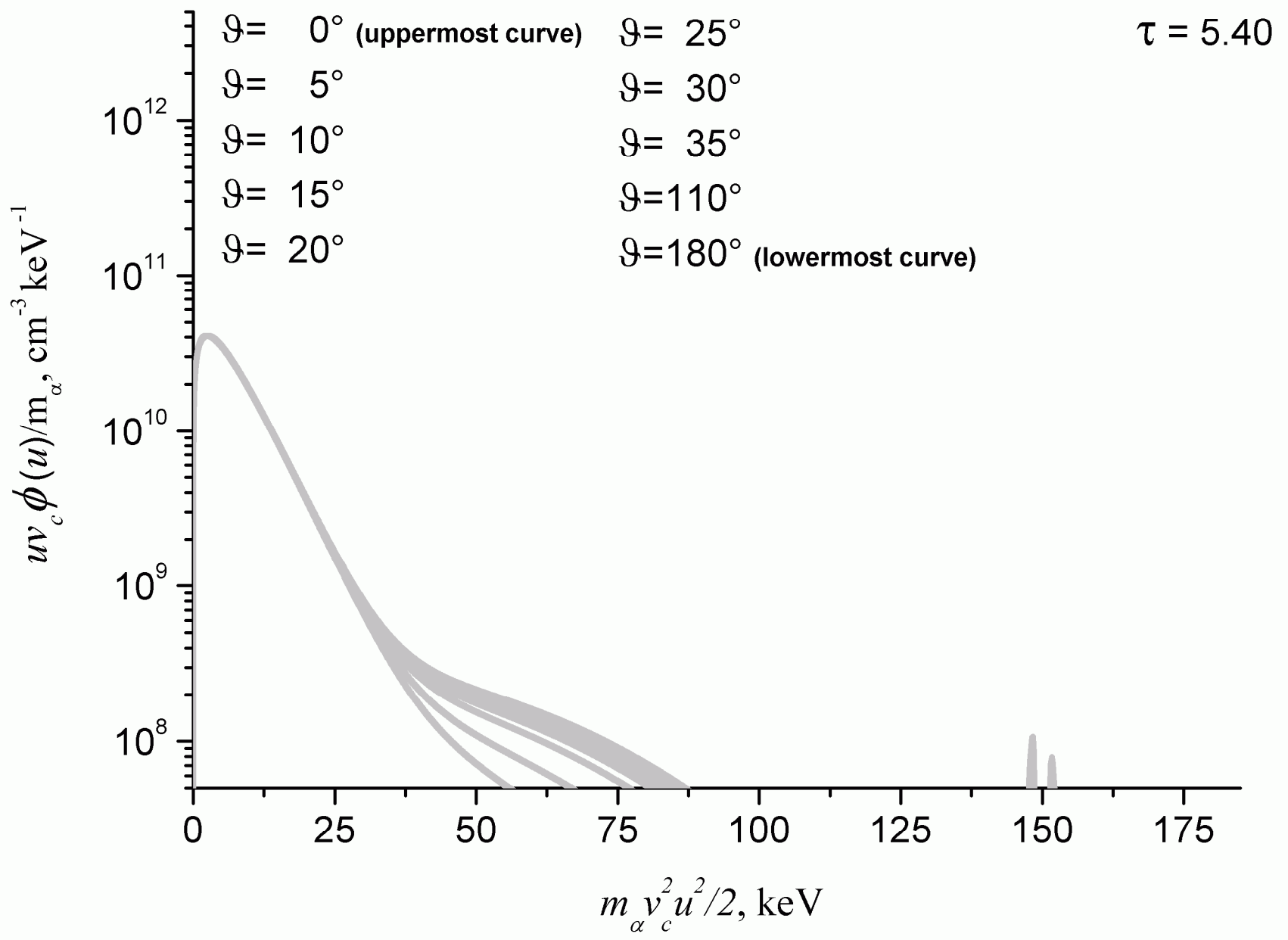

**(c)**

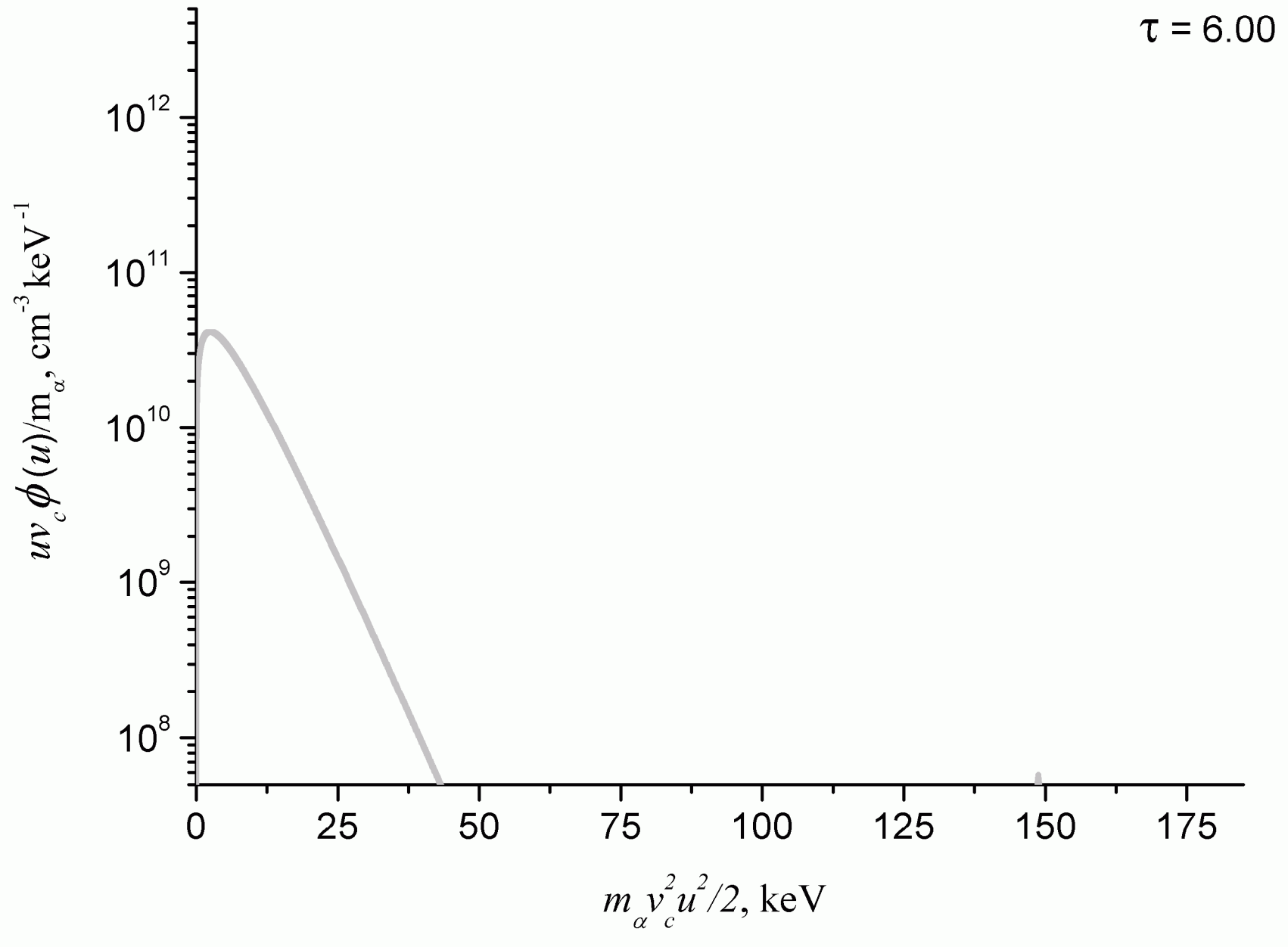

**(d)**

Fig. 4. Complete semi-analytical time-dependent solution of the anisotropic problem for 150 keV deuterons injected into isothermal $He_2^3 : He_2^4 : H_1^1$ (0.6 : 0.2 : 0.2) plasma at $T$ = 5 keV. Relaxation to statistical equilibrium after the source termination (a) for $\tau = 5.0095$, (b) for $\tau = 5.10$, (c) for $\tau = 5.40$, and (d) for $\tau = 6.00$.

Fig. 3 and Fig. 4 show calculation results for the parameters given in Table I and Table II as an example, assuming the source function (93) with $\tau_0 = 0.001$ and $\tau_1 = 5.0$ . The source is monodirectional in this example, and the injection angle is $\vartheta_0 = 0°$. Again, we show the energy distribution function $\frac{1}{m_\alpha}\sqrt{\frac{2E}{m_\alpha}}\phi(u,\zeta,\tau)$ versus $E = \frac{m_\alpha v_c^2 u^2}{2}$ instead of the solution $\phi(u,\zeta,\tau)$ itself. Numerical solutions of (108) were obtained for $u_L = 10^{-12}$, $u_R = 7$, and velocity grid dimension $N = 8192$. Time grid dimension was $M = 256 - 512$. Function $\phi(u,\zeta,\tau)$ was calculated by (95).

Fig. 3 illustrates the time evolution of the distribution function during the source action. At the beginning, Fig 3 (a), the distribution is peaked in the vicinity of the injection velocity and injection angle. After that test particles scatter in angle and slow down to lower energies, Fig 3 (b), and thermalize, Fig 3 (c), (d), approaching at low energies the statistical equilibrium with target plasma species. While the source with constant rate is acting, the population of thermalized particles is gradually increasing, and the high energy tail of the distribution, once developed, is not changing, as seen in Fig 3 (c) and (d), and corresponds to the high energy part of the steady state solution obtained in subsection 5.2. In our example at $\tau = 4.995$ test particle density is $n_\alpha = 0.535\times10^{13}$ cm$^{-3}$, which is about one order of magnitude lower than densities of target plasma species. We therefore consider matrices **A** and **B** as time-independent and neglect the collisions of test particles with themselves. If the source operation duration is longer, thermalized test particles can be taken into account as one of the components of target plasma.

Fig. 4 illustrates the relaxation of the test particle distribution to statistical equilibrium after the source termination. A gradual reduction in the high energy tail, Fig. 4 (a), (b), can be seen, followed by isotropisation, Fig. 4 (c), and Maxwellization, Fig. 4 (d).

Note that previously known time-dependent analytical solutions, such as [7], do not take into account velocity diffusion, and therefore are inapplicable to describe high-energy tails of the distributions correctly, and also cannot demonstrate thermalization. In addition, simplified solutions do not conserve the number of particles owing to the use of truncated Coulomb collision operators.

## 7. Summary

Semi-analytical stationary and nonstationary solutions of the kinetic equation with Coulomb collision term and a monoenergetic source function have been obtained, neglecting the spatial inhomogeneity and the electric field. Exact formulation of Coulomb collision term is used, assuming that azimuthal symmetry takes place and that the target plasma species are Maxwellian. The solutions reflect slowing-down, velocity diffusion, and pitch angle scattering effects.

Exact analytical isotropic steady state solution taking into account slowing-down and velocity diffusion, and exact analytical anisotropic steady state and time-dependent solutions taking into account slowing-down and pitch angle scattering have been obtained and used to verify the semi-analytical results. Previous simplified solutions, such as [6,7], are inappropriate to describe high energy tails of the distribution and are physically inadequate at lower energies, where the Maxwellization process should be observable.

The results may be useful in numerical modeling, especially concerning nuclear processes in magnetically confined plasmas, and also advanced localized, angle-resolved suprathermal particle diagnostic data analysis and interpretation.

## Acknowledgements

This work was partially supported by RFBR grant No. 09-02-13608-офи_ц, Rosatom Contract No. Н.4б.45.03.10.1011, Contract No. 02.740.11.0468 of Ministry of Education and Science of Russia and also Grant-in-Aid for JSPS Fellows No. 1806173 at National Institute for Fusion Science, Japan. The author appreciates discussions of applied aspects related to nuclear fusion neutron source development with Prof. B.V. Kuteev and Prof. V.Yu. Sergeev, and aspects related to our continuous collaboration on advanced fast particle diagnostics with Prof. S. Sudo, Dr. T. Ozaki, and Dr. N. Tamura. The author would also like to express his gratitude to Prof. Y.N. Dnestrovskii for a favourable discussion of the manuscript.